\documentstyle[preprint,aps,epsfig,here]{revtex}
\begin{document}
\title{Role of the  Delta (1232) in DIS on polarized $^3$He and extraction of the neutron spin structure function $g_{1}^{n}(x,Q^2)$}

\author{C. Boros$^{a}$, V. Guzey$^{a}$, M. Strikman$^{b}$, and A.W. Thomas$^{a,c}$}

\address{$^{a}$Special Research Centre for the Subatomic Structure of Matter (CSSM),\\
Adelaide University, Australia, 5005}

\address{$^{b}$The Pennsylvania State University, University Park, PA, USA}

\address{$^{c}$ The Department of Physics and Mathematical Physics, Adelaide University, Australia, 5005}

\preprint{
\vbox{
\hbox{ADP-00-30/T413}
}}

\maketitle

\begin{abstract}

We consider the effect of the transitions $n \rightarrow \Delta^{0}$ and $p \rightarrow \Delta^{+}$ in deep inelastic scattering on polarized $^3$He on the extraction of the neutron spin structure function $g_{1}^{n}(x,Q^2)$.
Making the natural assumption that these transitions are the dominant non-nucleonic contributions to the renormalization of the axial vector coupling constant in the $A=3$ system, 
we find that the effect of $\Delta$ increases $g_{1}^{n}(x,Q^2)$ by $10 \div 40$\% in the range $0.05 \le x \le 0.6$, where our considerations are applicable and most of the data for $g_{1}^{n}(x,Q^2)$ exist. 
\end{abstract}

\section{Introduction}
\label{sec:intro}

Deep inelastic scattering (DIS) of polarized leptons on polarized targets is used to study the spin structure functions of protons, neutrons, and light nuclei. The spin structure functions carry information about the distribution of the helicity of the target between its constituents.
Hence, studies of the spin structure functions are aimed at the understanding of the spin structure of nucleons and nuclei in terms of the underlying degrees of freedom, quarks and gluons.

This work is concerned with  
the neutron spin structure function $g_{1}^{n}(x,Q^2)$. Free neutron targets are not available. Instead, polarized deuterium and $^3$He targets are used as sources of information on the polarized neutron. Considerable experimental information on the structure function $g_{1}^{n}(x,Q^2)$ has been obtained so far. The HERMES collaboration at DESY \cite{HERMES} and the E154 experiment at SLAC \cite{E154} used a  polarized $^3$He target, while the  SMC collaboration at CERN \cite{SMC} and the E143 experiment at SLAC \cite{E143} used polarized deuterium. In both cases the extraction of the neutron structure function $g_{1}^{n}(x,Q^2)$ from the nuclear data required that nuclear effects be  taken into account.

The nuclear effects which play a role in polarized and unpolarized DIS on nuclei can be divided into coherent and incoherent contributions. Incoherent nuclear effects result from  the scattering of the incoming lepton on each individual nucleon, nucleon resonance, or virtual meson. 
They are present at all Bjorken $x$.

Coherent nuclear effects arise from the  interaction of the incoming lepton with two or more nucleons in the target. They  are typically concentrated at low values of Bjorken $x$.
Nuclear shadowing at $10^{-5} \div 10^{-4} \le x \le 0.05$ and antishadowing  at $0.05 \le x \le 0.2$ are examples of coherent effects. An analysis of the role of nuclear shadowing and antishadowing as well as the $\Delta \rightarrow N$ transitions
on the extraction of $g_{1}^{n}(x,Q^2)$ from the data on polarized DIS on $^3$He will be presented in a separate publication. 
In the present work, we do not consider shadowing and antishadowing effects since our main emphasis is on the region of high $x$.

For the case of polarized DIS, the major contribution comes from 
the incoherent scattering on the nucleons of the target. The nucleon-nucleon tensor force gives rise to sizable 
higher partial waves in  bound-state nuclear wave functions,
notably the $D$ wave in the deuteron ground-state wave function as well as the $S^{\prime}$ and $D$ waves in the $^3$He and $^3$H ground-state wave functions. The presence of these partial waves in the nuclear ground-state wave functions leads to spin depolarization (a decrease of the effective polarization) of the nucleons \cite{BW84}. In particular, in $^3$He the effective polarization of the neutron is often quoted as $86 \pm 2$\%, while the effective polarization of each proton is $-2.8 \pm 0.4$\% \cite{FGPBC}. These values represent the average of calculations with various nucleon-nucleon potentials and three-nucleon forces. However,
the large error bars are very conservative as they have been taken to be 3 times the average value of the spread of the calculated points about the fit to these points \cite{FGPBC}. In our analysis, we prefer to use the actual spread of the calculated values, namely,
 $P_{n}=86 \pm 0.8$\% and $P_{p}=-2.8 \pm 0.15$\% for the effective polarization of the neutron and the protons, respectively.

The importance  of  spin depolarization is well established. This effect was taken into account by the experimental collaborations named above when the neutron spin structure function $g_{1}^{n}(x,Q^2)$ was extracted from the DIS data on polarized deuterium and $^3$He. In order to extract the precise shape of $g_{1}^{n}(x,Q^2)$  one must also account for Fermi motion as well as binding and off-shell effects. 
For deuterium the calculations of Ref.\ \cite{MPT95} and for $^3$He those of  
Ref.\ \cite{CSPS} suggest that simply accounting for spin depolarization is quite a good approximation at $x \le 0.7$.

Until now, other incoherent nuclear effects such as nucleon resonances and meson-exchange currents have been  assumed to play a negligible role in polarized DIS. On the other hand, one knows that exchange currents involving the $\Delta$ resonance play a vital role in explaining the observed axial vector coupling constant of $^3$H.
Through the generalization of the Bjorken  sum rule to the tri-nucleon system one therefore knows that $\Delta$ must play a role in the spin structure functions of $^3$He and $^3$H. In this work we analyze the effect of $\Delta$ on the extraction of  $g_{1}^{n}(x,Q^2)$ from $g_{1}^{^3He}(x,Q^2)$.

\section{The role of $\Delta (1232)$ in polarized DIS on $^3$He}
\label{sec:one}

It is well known that exchange currents involving  a $\Delta$ isobar allow one  to account for the  4\% discrepancy
 between the experimental value  and theoretical predictions for the Gamow-Teller matrix element in triton beta decay \cite{Saito}. It was observed in Ref. \cite{FGS96} that this 4\% discrepancy  straightforwardly translates into the language of deep inelastic scattering on tri-nucleon systems. In particular, 
calculations with realistic bound-state wave functions of $^3$H and $^3$He
involving nucleons alone,
 underestimate the ratio of  the Bjorken sum rules for the tri-nucleon system and for the nucleon by the same amount.

The Bjorken sum rule was derived using the  algebra of currents \cite{Bjorken}. It reads    
\begin{equation}
\int_{0}^{1}\Big(g_{1}^{p}(x,Q^2)-g_{1}^{n}(x,Q^2)\Big)dx =\frac{1}{6}g_{A}\Big(1+O(\frac{\alpha_{s}}{\pi})\Big) \ ,
\label{bj}
\end{equation}
where $g_{A}$ is the axial vector coupling constant measured in the $\beta$ decay of neutron, $g_{A}=1.2670 \pm 0.0035$ \cite{Caso}. The QCD radiative corrections to this sum rule were calculated in perturbative QCD using the Operator Product Expansion. They are denoted by ``$O(\alpha_{s} / \pi)$''.

Analogously, one can write for the difference of the spin structure functions of $^3$H and $^3$He
\begin{equation}
\int_{0}^{1}\Big(g_{1}^{^3H}(x,Q^2)-g_{1}^{^3He}(x,Q^2)\Big)dx =\frac{1}{6}g_{A}|_{triton}\Big(1+O(\frac{\alpha_{s}}{\pi})\Big) \ ,
\label{bj2}
\end{equation}  
where $g_{A}|_{triton}$ is  the axial vector coupling constant measured in the $\beta$ decay of triton, $g_{A}|_{triton}=1.211 \pm 0.002$ \cite{Budick}.

Taking the ratio of Eqs.\ (\ref{bj}) and (\ref{bj2}), one obtains
\begin{equation}
\eta \equiv \frac{g_{A}|_{triton}}{g_{A}}=\frac{\int_{0}^{1}\Big(g_{1}^{^3H}(x,Q^2)-g_{1}^{^3He}(x,Q^2)\Big)dx}{\int_{0}^{1}\Big(g_{1}^{p}(x,Q^2)-g_{1}^{n}(x,Q^2)\Big)dx}=0.956 \pm 0.004 \ .
\label{master}
\end{equation}
Note that in Eq.\ (\ref{master}) the QCD radiative corrections cancel exactly.

On the other hand, calculations with exact wave functions of the tri-nucleon system involving only nucleons predict that 
\begin{eqnarray}
&&\int \Big(g_{1}^{^3H}(x,Q^2)-g_{1}^{^3He}(x,Q^2)\Big)dx= \nonumber\\
&&(1-\frac{4}{3}P_{S^{\prime}}-\frac{2}{3}P_{D})\int \Big( (g_{1}^{p}(x,Q^2)-g_{1}^{n}(x,Q^2)\Big)dx \equiv \nonumber\\
&&(P_{n}-2P_{p}) \int  \Big(g_{1}^{p}(x,Q^2)-g_{1}^{n}(x,Q^2)\Big)dx \ ,
\label{chain}
\end{eqnarray}
where $P_{S^{\prime}}$ and $P_{D}$ are the probabilities of the corresponding partial waves in the bound-state wave function of $^3$He. Using $P_{n}=0.86 \pm 0.008$ and $2P_{p}=-0.056 \pm 0.003$
 and substituting Eq.\ (\ref{chain}) into Eq.\ (\ref{master}) one obtains 
\begin{equation}
\frac{\int_{0}^{1}\Big(g_{1}^{^3H}(x,Q^2)-g_{1}^{^3He}(x,Q^2)\Big)dx}{\int_{0}^{1}\Big(g_{1}^{p}(x,Q^2)-g_{1}^{n}(x,Q^2)\Big)dx}=0.916 \pm 0.009 \ .
\label{th}
\end{equation}
In Eq.\ (\ref{th}) the errors for $P_{n}$ and $2P_{p}$ are not correlated and therefore have been added in quadrature.

By comparing Eqs.\ (\ref{master}) and (\ref{th}) one can see that calculations of the structure functions  $g_{1}^{^3H}(x,Q^2)$ and $g_{1}^{^3He}(x,Q^2)$ based on nucleons alone underestimate the quenching factor $\eta$ in Eq.\ (\ref{master}) by about $4$\%.

As explained above, it is natural to expect that this discrepancy can be accounted for by including  the DIS diagrams which correspond to the transitions $n \rightarrow \Delta^{0}$ and $p \rightarrow \Delta^{+}$. This corresponds precisely  to the two-body exchange currents involving the $\Delta$ which appear in calculations of the Gamow-Teller matrix element for the triton beta decay. Fig.\ \ref{fig:one} demonstrates the correspondence between the interference diagrams in polarized DIS and the two-body exchange currents involving a  $\Delta$ entering the  Gamow-Teller matrix element   calculations.
 
In Ref.\ \cite{Saito} was shown that the contribution of the diagrams of Fig.\ \ref{fig:one} increases 
the theoretical prediction of  the axial vector  coupling constant of the triton by 4\% and makes it consistent with the experimental value.
 By analogy, we  assume that the contribution of the interference terms, 
$n \rightarrow \Delta^{0}$ and $p \rightarrow \Delta^{+}$,
in polarized DIS
will make the theoretical prediction of  \ Eq.\ (\ref{th}) equal to the experimental value of Eq.\ (\ref{master}).

Note also  that the contribution of the DIS diagrams involving the transition 
$\Delta \rightarrow \Delta$ 
 is neglected since the contribution of the corresponding diagrams to triton beta decay is negligible.

Taking into account the interference terms in polarized DIS on tri-nucleon systems, the spin structure functions of $^3$He and $^3$H can be expressed as
\begin{eqnarray}
&&g_{1}^{^3He}(x,Q^2)=\int_{x}^{A} \frac{dy}{y} \Delta f_{n/^3He}(y)g_{1}^{n}(x/y,Q^2)+\int_{x}^{A} \frac{dy}{y} \Delta f_{p/^3He}(y)g_{1}^{p}(x/y,Q^2)+ \nonumber\\
&&+\int_{x}^{A} \frac{dy}{y} \Delta f_{n \rightarrow \Delta^{0}/^3He}(y) g_{1}^{n \rightarrow \Delta^{0}}(x/y,Q^2)+\int_{x}^{A} \frac{dy}{y}\Delta f_{p \rightarrow \Delta^{+}/^3He}(y) g_{1}^{p \rightarrow \Delta^{+}}(x/y,Q^2) \ , \nonumber \\
&&g_{1}^{^3H}(x,Q^2)=\int_{x}^{A} \frac{dy}{y}\Delta f_{n/^3H}(y)g_{1}^{n}(x/y,Q^2)+\int_{x}^{A} \frac{dy}{y} \Delta f_{p/^3H}(y)g_{1}^{p}(x/y,Q^2)+ \nonumber\\
&&+\int_{x}^{A} \frac{dy}{y}\Delta f_{n \rightarrow \Delta^{0}/^3H}(y) g_{1}^{n \rightarrow \Delta^{0}}(x/y,Q^2)+\int_{x}^{A} \frac{dy}{y}\Delta  f_{p \rightarrow \Delta^{+}/^3H}(y) g_{1}^{p \rightarrow \Delta^{+}}(x/y,Q^2) \ ,
\label{def}
\end{eqnarray}
where $\Delta f_{n/^3He}(y)$ ($\Delta f_{n/^3H}(y)$), $\Delta f_{p/^3He}(y)$ ($\Delta f_{p/^3H}(y)$), $\Delta f_{n \rightarrow \Delta^{0}/^3He}(y)$ ($\Delta f_{n \rightarrow \Delta^{0}/^3H}(y)$), and $\Delta f_{p \rightarrow \Delta^{+}/^3He}(y)$ ($\Delta f_{p \rightarrow \Delta^{+}/^3H}(y)$) are the spin-dependent light-cone momentum distributions of the neutron, proton, $n \rightarrow \Delta^{0}$,  and $p \rightarrow \Delta^{+}$ interference terms in $^3$He  ($^3$H), respectively; $g_{1}^{n \rightarrow \Delta^{0}}(x,Q^2)$ and $g_{1}^{p \rightarrow \Delta^{+}}(x,Q^2)$ are the spin structure functions of the corresponding interference terms.

In the approximation where Fermi motion and off-shell effects are negligible so $\Delta f_{i/^3He}(y) \propto \delta(y-1)$ ($\Delta f_{i/^3H}(y) \propto \delta(y-1)$) we find 

\begin{eqnarray}
&&g_{1}^{^3He}(x,Q^2)=P_{n}g_{1}^{n}(x,Q^2)+2P_{p}g_{1}^{p}(x,Q^2) \nonumber\\
&&+2P_{n \rightarrow \Delta^{0}} g_{1}^{n \rightarrow \Delta^{0}}(x,Q^2)+4P_{p \rightarrow \Delta^{+}} g_{1}^{p \rightarrow \Delta^{+}}(x,Q^2) \ , \nonumber \\
&&g_{1}^{^3H}(x,Q^2)=P_{n}g_{1}^{p}(x,Q^2)+2P_{p}g_{1}^{n}(x,Q^2) \nonumber\\
&&-2P_{n \rightarrow \Delta^{0}} g_{1}^{p \rightarrow \Delta^{+}}(x,Q^2)-4P_{p \rightarrow \Delta^{+}} g_{1}^{n \rightarrow \Delta^{0}}(x,Q^2) \ .
\label{defb}
\end{eqnarray}
In Eq.\ (\ref{defb}) $P_{n \rightarrow \Delta^{0}}$ and $P_{p \rightarrow \Delta^{+}}$ stand for the effective polarization of the interference transitions $n \rightarrow \Delta^{0}$ and $p \rightarrow \Delta^{+}$ in $^3$He, respectively. Additional factors of two in front of the interference terms correspond to the sum of the $N \rightarrow \Delta$ and $\Delta \rightarrow N$ transitions. 
The minus sign in front of the interference term contribution to $g_{1}^{^3H}(x,Q^2)$ originates due to the sign convention 
\begin{eqnarray}
&&P_{n \rightarrow \Delta^{0}} \equiv  P_{n \rightarrow \Delta^{0}/^3He}= -P_{p \rightarrow \Delta^{+}/^3H} \ , \nonumber\\
&&P_{p \rightarrow \Delta^{+}} \equiv  P_{p \rightarrow \Delta^{+}/^3He}= -P_{n \rightarrow \Delta^{0}/^3H} \ .
\end{eqnarray}

In the next section we shall make predictions for spin structure functions $g_{1}^{n \rightarrow \Delta^{0}}(x,Q^2)$ and  $g_{1}^{p \rightarrow \Delta^{+}}(x,Q^2)$ using the quark model and estimate the effective polarizations $P_{n \rightarrow \Delta^{0}}$ and $P_{p \rightarrow \Delta^{+}}$ using Eq.\ (\ref{master}) as a guide.    

\section{Interference spin structure functions and the effective polarizations}
\label{sec:two}

The contribution to the nuclear  spin structure functions associated with  the  $\Delta \rightarrow N$ and $N \rightarrow \Delta$ interference    terms, $g_{1}^{n \rightarrow \Delta^{0}}(x,Q^2)$ and  $g_{1}^{p \rightarrow \Delta^{+}}(x,Q^2)$,  can be calculated within the framework of the valence quark  model. Since only valence quarks are present in this picture, the model, and hence, our predictions for the interference spin structure functions $g_{1}^{n \rightarrow \Delta^{0}}(x,Q^2)$ and  $g_{1}^{p \rightarrow \Delta^{+}}(x,Q^2)$
 are valid where the valence parton picture is applicable. The following simple analysis shows that polarized valence quarks dominate over polarized sea quarks  at $Q^2 \le 4$ GeV$^2$ and not very low $x$. 
Using the parameterization of spin-dependent parton distributions from Ref. \cite{GRSV}, one can readily check that  polarized distribution for valence $u$ and $d$ quarks are larger than the polarized sea quark  distributions in the range $0.34 \le Q^2 \le 4$ GeV$^2$ and $0.05 \le x \le 0.8$ by at least a factor of five. This justifies the use of the valence (constituent) quark picture at these values of  $Q^2$ and $x$.

The probabilities to find polarized {\it up}, {\it down}, and {\it strange} quarks for the octet of baryons, the decuplet of baryon resonances and the interference between octet and decuplet states were derived in Ref. \cite{BT}, using SU(6) wave functions with energies perturbed by the standard spin-dependent  
hyperfine interactions \cite{CT88} -- following earlier work for the nucleon \cite{ST}. 

The spin structure function $g_{1}(x,Q^2)$ for the proton and  neutron, and the interference terms, $p \rightarrow \Delta^{+}$ and $n \rightarrow \Delta^{0}$,  
within the framework of the valence quark parton model is defined as
\begin{equation}
g_{1}(x,Q^2)=\frac{1}{2}\Big(\frac{4}{9}\Delta u(x,Q^2) + \frac{1}{9}\Delta d(x,Q^2)\Big) \ ,
\label{g1}
\end{equation}
where $\Delta u(x,Q^2) \equiv u^{\uparrow}(x,Q^2)-u^{\downarrow}(x,Q^2)$ and $\Delta d(x,Q^2) \equiv d^{\uparrow}(x,Q^2)-d^{\downarrow}(x,Q^2)$; $u^{\uparrow}(x,Q^2)$ ($d^{\uparrow}(x,Q^2)$) and $u^{\downarrow}(x,Q^2)$ ($d^{\downarrow}(x,Q^2)$) are the probabilities to find the up (down) quark with helicity parallel and antiparallel, respectively, to the helicity of the target.

Using the results of Ref.\ \cite{BT} and the definition (\ref{g1}) one can express the spin structure functions for the proton and neutron as well as the interference terms, $p \rightarrow \Delta^{+}$ and $n \rightarrow \Delta^{0}$, at some initial scale $Q_{0}$ as
\begin{eqnarray}
&&g_{1}^{p}(x,Q_{0}^2)=\frac{1}{18}\Big(6G_s(x,Q_{0}^2)-G_v(x,Q_{0}^2)\Big) \ , \nonumber\\
&&g_{1}^{n}(x,Q_{0}^2)=\frac{1}{12}\Big(G_s(x,Q_{0}^2)-G_v(x,Q_{0}^2)\Big) \ , \nonumber\\
&&g_{1}^{p \rightarrow \Delta^{+}}(x,Q_{0}^2)=\frac{\sqrt{2}}{9}G_v(x,Q_{0}^2) \ , \nonumber\\
&&g_{1}^{n \rightarrow \Delta^{0}}(x,Q_{0}^2)=\frac{\sqrt{2}}{9}G_v(x,Q_{0}^2) \ ,
\label{m2}
\end{eqnarray}
where $G_s(x,Q_{0}^2)$ and $G_v(x,Q_{0}^2)$ are the contributions to the spin structure functions associated with 
a pair of spectator quarks in the
baryon wave function with
 $S=0, I=0$ and $S=1, I=1$, respectively. These contributions  can be calculated using, for example, the MIT Bag model with $Q_{0}^2$=0.23 GeV$^2$, as was done in Ref.\ \cite{BT}.

One can also derive a sum rule, analogous to the Bjorken sum rule, which relates in a model-independent way the sum of the first moments of $g_{1}^{p \rightarrow \Delta^{+}}(x,Q^2)$ and $g_{1}^{n \rightarrow \Delta^{0}}(x,Q^2)$ to a certain axial current matrix element.
The derivation is completely analogous to the one for the Bjorken sum rule. That is, 
the operator product expansion or algebra of currents relates the commutator of two electromagnetic currents, whose matrix element defines the usual 
hadronic electromagnetic tensor $W_{\mu \nu}$ of DIS,
 to the axial current. Sandwiching this commutator between baryon states and using an  unsubtracted dispersion relation for the structure function $g_{1}(x,Q^2)$ constrains the integral $\int g_{1}(x,Q^2) dx$. If we choose the initial state to be a nucleon and the final state to be a $\Delta (1232)$ resonance we arrive at the following relationships
\begin{eqnarray}
&&\int_{0}^{1} g_{1}^{p \rightarrow \Delta^{+}}(x,Q^2)dx=\frac{1}{2} B_{1}^{p \rightarrow \Delta^{+}}\Big(1+O(\frac{\alpha_{s}}{\pi})\Big)  \ , \nonumber\\
&&\int_{0}^{1}  g_{1}^{n \rightarrow \Delta^{0}}(x,Q^2)dx =\frac{1}{2}B_{1}^{n \rightarrow \Delta^{0}}\Big(1+O(\frac{\alpha_{s}}{\pi})\Big)  \ .
\label{s1}
\end{eqnarray}
The matrix elements $B_{1}$ are defined as
\begin{eqnarray}
&&2\, s^{\mu} B_{1}^{p \rightarrow \Delta^{+}} \equiv \langle \Delta^{+},s|\frac{4}{9} \bar{u}\gamma^{\mu} \gamma_{5} u +\frac{1}{9} \bar{d}\gamma^{\mu} \gamma_{5} d|p,s \rangle \ , \nonumber\\
&&2\, s^{\mu} B_{1}^{n \rightarrow \Delta^{0}} \equiv \langle \Delta^{0},s|\frac{4}{9} \bar{u}\gamma^{\mu} \gamma_{5} u +\frac{1}{9} \bar{d}\gamma^{\mu} \gamma_{5} d|n,s \rangle \ ,
\label{sf2}
\end{eqnarray}
where  $s^{\mu}$ is the polarization vector of the nucleon and $\Delta$ defined as in Ref.\ \cite{Manohar92}. 

Using the representation of the wave functions of the proton, neutron, and $\Delta$ in terms of quark fields and the standard commutation relationships between the quark fields 
 one can relate the sum of the matrix elements $B_{1}^{p \rightarrow \Delta^{+}}$ and $B_{1}^{n \rightarrow \Delta^{0}}$ to the axial vector coupling constant for the beta decay $\nu_{\mu} \,p \rightarrow \mu^{-}\,\Delta^{++}$. Decomposing the electromagnetic current in Eq.\ (\ref{sf2}) into isovector and isoscalar components one can readily obtain
\begin{eqnarray}
&&\langle \Delta^{+},s|\bar{u}\gamma^{\mu} \gamma_{5} u -\bar{d}\gamma^{\mu} \gamma_{5} d|p,s \rangle=-2\langle \Delta^{+},s|\bar{u}\gamma^{\mu} \gamma_{5} d |n,s \rangle \ , \nonumber\\
&&\langle \Delta^{0},s|\bar{u}\gamma^{\mu} \gamma_{5} u -\bar{d}\gamma^{\mu} \gamma_{5} d|n,s \rangle=-2\langle \Delta^{+},s|\bar{u}\gamma^{\mu} \gamma_{5} d |n,s \rangle \ .
\label{sf3}
\end{eqnarray}
It also follows from Ref.\ \cite{BT} that
\begin{eqnarray}
&&\langle \Delta^{+},s|\bar{u}\gamma^{\mu} \gamma_{5} u +\bar{d}\gamma^{\mu} \gamma_{5} d|p,s \rangle=0 \ , \nonumber\\
&&\langle \Delta^{0},s|\bar{u}\gamma^{\mu} \gamma_{5} u +\bar{d}\gamma^{\mu} \gamma_{5} d|n,s \rangle=0 \ .
\label{sf4}
\end{eqnarray}
Using Eqs.\ (\ref{sf2}), (\ref{sf3}), and (\ref{sf4}) one can write
\begin{equation}
2\, s^{\mu}(B_{1}^{p \rightarrow \Delta^{+}}+B_{1}^{n \rightarrow \Delta^{0}})=-
\frac{2}{3} \langle \Delta^{+},s|\bar{u}\gamma^{\mu} \gamma_{5} d |n,s \rangle \ .
\label{sf4a}
\end{equation}

The commutation relationships between the quark fields
relate 
the latter matrix element to  the effective axial  vector  coupling constant in the reaction $\nu_{\mu} \,p \rightarrow \mu^{-}\,\Delta^{++}$, $g_{A}(p \rightarrow \Delta^{++})$:
\begin{equation}
\langle \Delta^{+},s|\bar{u}\gamma^{\mu} \gamma_{5} d |n,s \rangle =-\frac{1}{\sqrt{3}} \langle \Delta^{++},s|\bar{u}\gamma^{\mu} \gamma_{5} d |p,s \rangle \equiv -\frac{1}{\sqrt{3}}2\,s^{\mu} g_{A}(p \rightarrow \Delta^{++}) \ .
\label{sf5}
\end{equation}
Combining Eqs.\ (\ref{sf4a}), (\ref{sf5}), and (\ref{s1}) one obtains the following sum rule
\begin{equation}
\int_{0}^{1}\Big(g_{1}^{p \rightarrow \Delta^{+}}(x,Q^2)+g_{1}^{n \rightarrow \Delta^{0}}(x,Q^2)\Big)dx=\frac{1}{3\sqrt{3}}g_{A}(p \rightarrow \Delta^{++})\Big(1+O(\frac{\alpha_{s}}{\pi})\Big) \ .
\label{sr1}
\end{equation}

The sum rule (\ref{sr1}) is exact (modulo the QCD radiative corrections) in the limit of SU(6) symmetry. Indeed, in this case  $\int_{0}^{1}G_{v}(x,Q_{0}^2)dx=1$ and the left hand side of Eq.\ (\ref{sr1}) equals $2\sqrt{2}/9$. On the other hand, $g_{A}(p \rightarrow \Delta^{++})=2\sqrt{2/3}$ \cite{Alfaro} and the right hand side of Eq.\ (\ref{sr1}) equals $2\sqrt{2}/9$ too. 
  
We also expect that SU(6)  should be a good approximation for the right hand side of Eq.\ (\ref{sr1}) because it works qualitatively well for the form factor $C_{5}^{A}(0)$ associated with the reaction
$\nu_{\mu} \, p \rightarrow \mu^{-} \Delta^{++}$. The theoretical analysis of Ref.\ \cite{Alvarez} of the experimental data from the BNL experiment \cite{Kitagaki} found that
  $C_{5}^{A}(0)=1.22 \pm 0.06$. This value is close to the SU(6) prediction, $C_{5}^{A}(0)=1.15$.

In order to successfully apply Eq.\ (\ref{defb}) to extract the neutron spin structure function $g_{1}^{n}(x,Q^2)$ from the $^3$He data
 one needs to address two issues: the effective polarizations of the interference terms, $P_{n \rightarrow \Delta^{0}}$ and $P_{p \rightarrow \Delta^{+}}$
 and the $Q^2$ and $x$ dependence of the spin structure functions of the interference terms, $g_{1}^{n \rightarrow \Delta^{0}}(x,Q^2)$ and $g_{1}^{p \rightarrow \Delta^{+}}(x,Q^2)$.

 The first question can be readily answered using the sum rule (\ref{sr1}). Substituting Eqs.\ (\ref{defb}) into Eq.\ (\ref{master}) and using Eq.\ (\ref{th})  one obtains
\begin{equation}
0.956=0.916+2\,(P_{n \rightarrow \Delta^{0}}+2P_{p \rightarrow \Delta^{+}})
\frac{\int_{0}^{1}dx(g_{1}^{n \rightarrow \Delta^{0}}(x,Q^2)+g_{1}^{p \rightarrow \Delta^{+}}(x,Q^2))}{\int_{0}^{1}dx(g_{1}^{n}(x,Q^2)-g_{1}^{p}(x,Q^2))} \ .
\label{sr2}
\end{equation}
We have assumed that the QCD radiative corrections in Eq.\ (\ref{sr2}) cancel.

Since we have assumed that the contributions of the $n \rightarrow \Delta^{0}$ and $p \rightarrow \Delta^{+}$ transitions make the theoretical prediction and the 
experimental value of 
  $g_{A}|_{triton}$ consistent,
the effective polarizations of the interference terms,  $P_{n \rightarrow \Delta^{0}}$ and $P_{p \rightarrow \Delta^{+}}$, should be such that  Eq.\ (\ref{sr2}) is satisfied. Thus, using the sum rule (\ref{sr1}) and Eq.\ (\ref{sr2}) one can write 
\begin{equation}
2(P_{n \rightarrow \Delta^{0}}+2P_{p \rightarrow \Delta^{+}})=-\sqrt{3} \times 0.04\frac{g_{A}}{g_{A}(p \rightarrow \Delta^{++})}=-0.027  \ .
\label{sr3}
\end{equation}
In Eq.\ (\ref{sr3}) we have assumed the SU(6) value for $g_{A}(p \rightarrow \Delta^{++})$, namely $g_{A}(p \rightarrow \Delta^{++})=2\sqrt{2/3}$.

Therefore, the spin structure function $g_{1}^{^3He}(x,Q^2)$  of Eq.\ (\ref{defb})  can be written as
\begin{equation}
g_{1}^{^3He}(x,Q^2)=P_{n}g_{1}^{n}(x,Q^2)+2P_{p}g_{1}^{p}(x,Q^2)-0.027\,g_{1}^{n \rightarrow \Delta^{0}}(x,Q^2) \ .
\label{sf10}
\end{equation}
We  stress that Eq.\ (\ref{sf10}) is an approximation which neglects Fermi motion and off-shell effects.

Next we need to know the interference structure function,  $g_{1}^{n \rightarrow \Delta^{0}}(x,Q^2)$, as a function of $Q^2$ and $x$.
In Ref.\ \cite{BT} $G_s(x,Q^2)$ and $G_v(x,Q^2)$ which enter Eq.\ (\ref{m2}) were evaluated within the framework of the MIT Bag model. 
 Then, in order to make a comparison to other parameterizations of polarized quark densities and to spin structure functions,  the QCD evolution from the bag scale $Q_{0}^2=0.23$ GeV$^2$ to large $Q^2$ was performed. 

Instead of using a particular model for $G_s(x,Q^2)$ and $G_v(x,Q^2)$ and then performing the QCD evolution, one can relate $g_{1}^{n \rightarrow \Delta^{0}}(x,Q^2)$ to  $g_{1}^{p}(x,Q^2)$ and  $g_{1}^{n}(x,Q^2)$ in a model-independent way.
Using Eq.\ (\ref{m2}) one can write
\begin{equation}
g_{1}^{p \rightarrow \Delta^{+}}(x,Q^2)=g_{1}^{n \rightarrow \Delta^{0}}(x,Q^2)=\frac{2\sqrt{2}}{5}\Big(g_{1}^{p}(x,Q^2)-4g_{1}^{n}(x,Q^2)\Big) \ .
\label{m3}
\end{equation}
Since Eq.\ (\ref{m3}) is based solely on the Clebsh-Gordon coefficients used to construct the nucleon and $\Delta$ wave functions, it holds regardless 
of whether the
SU(6) symmetry of the baryon wave functions is broken or not. Eq.\ (\ref{m3})
 should be valid at all $x$ and $Q^2$ where the picture of nucleons and $\Delta$'s being composed of valence quarks holds.  No information about the dynamics of valence quarks, or information on $G_s(x,Q^2)$ and $G_v(x,Q^2)$, is needed.

Using  Eq.\ (\ref{m3}) in Eq.\ (\ref{sf10}) one can write the following master equation for the spin structure function $g_{1}^{^3He}(x,Q^2)$
\begin{equation}
g_{1}^{^3He}(x,Q^2)=P_{n}g_{1}^{n}(x,Q^2)+2P_{p}g_{1}^{p}(x,Q^2)-0.027\,\frac{2\sqrt{2}}{5}\Big(g_{1}^{p}(x,Q^2)-4g_{1}^{n}(x,Q^2)\Big)  \ .
\label{sf22}
\end{equation}
Eq.\ (\ref{sf22}) describes $g_{1}^{^3He}(x,Q^2)$ as a sum of the contributions  from  the effective polarizations of the neutron and proton and the contribution of the interference terms $N \rightarrow \Delta$ and $\Delta \rightarrow N$. 
As  explained above, Eq.\ (\ref{sf22}) neglects  Fermi motion and off-shell effects, which is expected to be  a good approximation for $x \le 0.7$.

\section{Extraction of \lowercase{$g_{1}^{n}(x,\uppercase{Q}^2)$} from $^3$He data}
\label{sec:three}
 
Eq.\ (\ref{sf22}) can be used estimate the role played by a $\Delta$ in extracting the neutron spin structure function $g_{1}^{n}(x,Q^2)$ from the DIS data taken on a polarized $^3$He target.

Let us denote $g_{1exp.}^{n}(x,Q^2)$ the neutron spin structure function obtained from  Eq.\ (\ref{sf22}) when the contribution of the $\Delta$ is omitted
\begin{equation}
g_{1}^{^3He}(x,Q^2)=P_{n}g_{1exp.}^{n}(x,Q^2)+2P_{p}g_{1}^{p}(x,Q^2) \ .
\label{sf33}
\end{equation}
Combining Eqs.\ (\ref{sf22}) and (\ref{sf33}) we find the relationship between the  theoretical prediction for $g_{1}^{n}(x,Q^2)$ when the effect of the  $\Delta$ is present and  $g_{1exp.}^{n}(x,Q^2)$
\begin{eqnarray}
g_{1}^{n}(x,Q^2)&&=\Big(g_{1exp.}^{n}(x,Q^2)+\frac{0.027}{P_{n}}\frac{2\sqrt{2}}{5}g_{1}^{p}(x,Q^2)\Big)\times\Big(1+\frac{0.027}{P_{n}}\frac{8\sqrt{2}}{5})^{-1}= \nonumber\\
&&0.934\,\Big(g_{1exp.}^{n}(x,Q^2)+0.0178\,g_{1}^{p}(x,Q^2)\Big) \ .
\label{master2}
\end{eqnarray}
One can also  represent the result of Eq.\ (\ref{master2}) in the form of the ratio $g_{1}^{n}(x,Q^2)/g_{1exp.}^{n}(x,Q^2)$
\begin{equation}
\frac{g_{1}^{n}(x,Q^2)}{g_{1exp.}^{n}(x,Q^2)}=0.934+0.0178\frac{g_{1}^{p}(x,Q^2)}{g_{1exp.}^{n}(x,Q^2)} \ .
\label{master3}
\end{equation}

\begin{figure}
\hspace{-2cm}
\includegraphics[scale=.9]{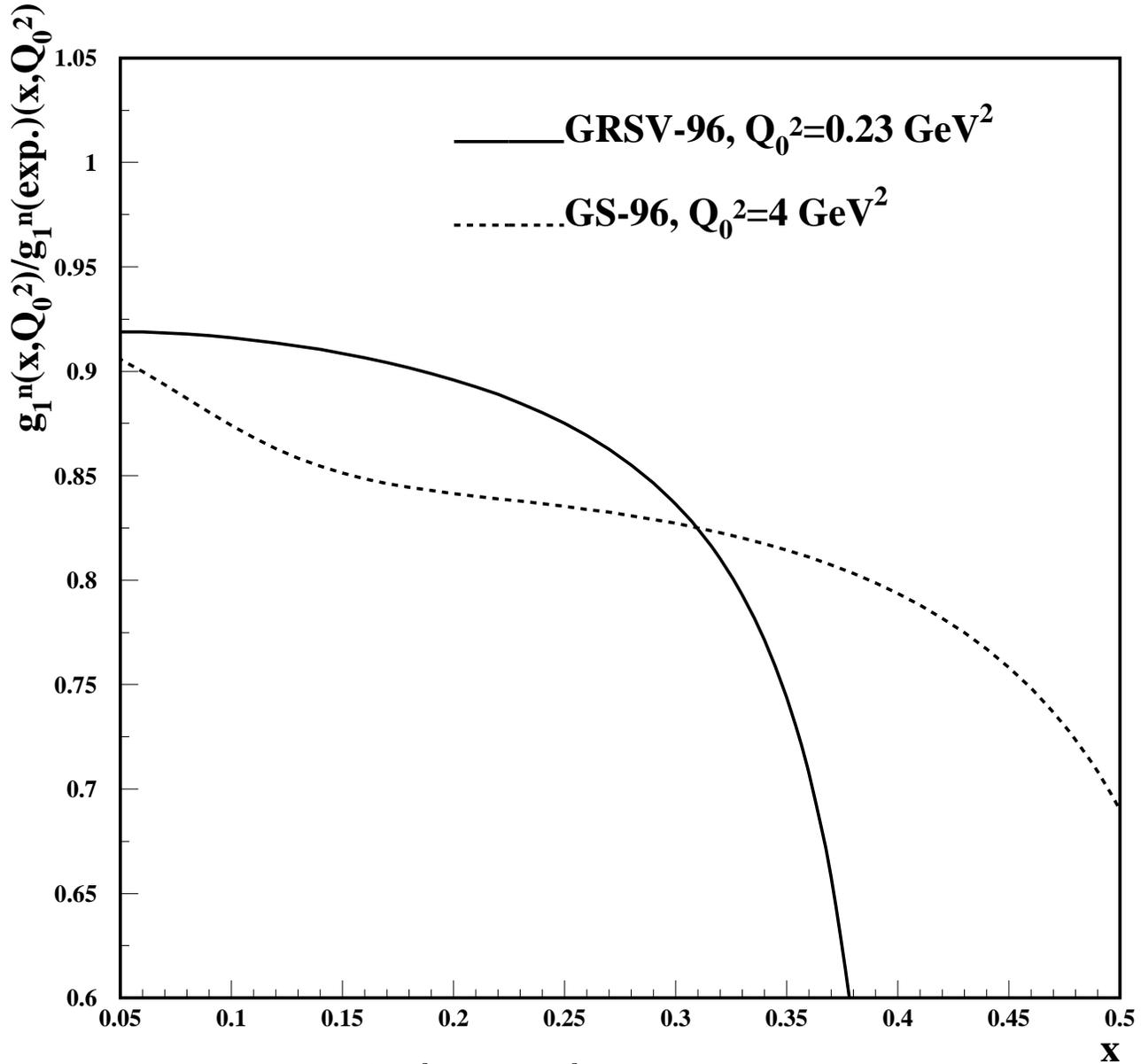}
\caption{The ratio $g_{1}^{n}(x,Q^2)/g_{1exp.}^{n}(x,Q^2)$ of Eq.\ (\ref{master3})  for two different parameterizations of the spin-dependent parton densities at the initial scales $Q_{0}^2$=0.23 GeV$^2$ and $Q_{0}^2$=4 GeV$^2$, respectively. This ratio would be unity if the effect of the $\Delta$ were unimportant.}
\label{fig:two}
\end{figure}

In order to demonstrate the magnitude of the nuclear effect associated with the $\Delta$ we show in Fig.\ \ref{fig:two} the ratio $g_{1}^{n}(x,Q^2)/g_{1exp.}^{n}(x,Q^2)$ of Eq.\ (\ref{master3}) as a function of $x$. The solid line corresponds to the parameterization of spin-dependent parton densities of Ref.\ \cite{GRSV} at the initial scale $Q_{0}^2$=0.23 GeV$^2$. The dotted line corresponds to  the parameterization of Ref. \cite{GS96} at a higher initial scale $Q_{0}^2$=4 GeV$^2$. The considered ratio would be  unity if the effect of the $\Delta$ were unimportant. However, from Fig.\ \ref{fig:two} one can see that the contributions of the interference terms, $N \rightarrow \Delta$ and  $\Delta \rightarrow N$, do modify the neutron spin structure function $g_{1exp.}^{n}(x,Q^2)$ -- it is reduced by approximately 7\% in the range $0.05 \le x \le 0.1 \div 0.2$ and drops significantly  for $x > 0.1 \div 0.2$
 when the parameterization of Ref.\ \cite{GRSV} is used. The tendency is similar for  the parameterization of Ref.\ \cite{GS96}.

Using Eq.\ (\ref{master2}) one can also estimate the change in the first moment of $g_{1}^{n}(x,Q^2)$. Assuming that the effects associated with the $\Delta$ are present at $0.05 \le x \le 0.8$ in Eq.\ (\ref{master2}) one finds
\begin{eqnarray}
&&\int_{0}^{1} g_{1exp.}^{n}(x,Q_{0}^2)dx=-0.0621 \ , \nonumber\\
&&\int_{0}^{1} g_{1}^{n}(x,Q_{0}^2)dx=-0.0574 
\label{mom1}
\end{eqnarray}
for the parameterization of Ref.\  \cite{GRSV} at $Q_{0}^2$=0.23 GeV$^2$, and 
\begin{eqnarray}
&&\int_{0}^{1} g_{1exp.}^{n}(x,Q_{0}^2)dx=-0.0574 \ , \nonumber\\
&&\int_{0}^{1} g_{1}^{n}(x,Q_{0}^2)dx=-0.0540
\label{mom2}
\end{eqnarray}
for the parameterization of Ref.\  \cite{GS96} at $Q_{0}^2$=4 GeV$^2$. For the two  parameterizations of spin-dependent parton densities considered here, the contribution of the interference $N \rightarrow \Delta$ and  $\Delta \rightarrow N$ terms changes the first moment of $g_{1}^{n}(x,Q^2)$  by 8\% and 6\%, respectively.

\begin{figure}
\hspace{-2cm}
\includegraphics[scale=.9]{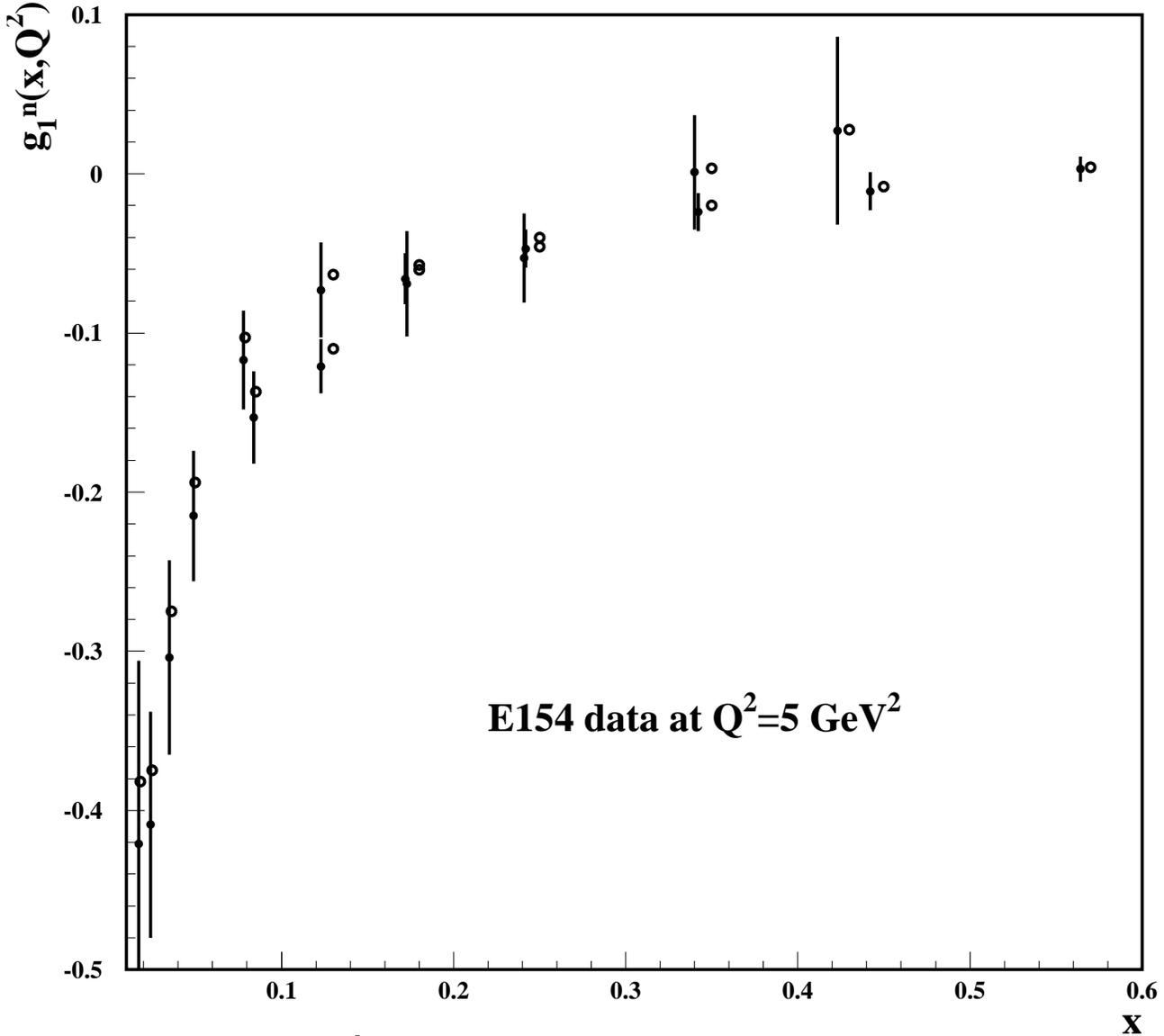}
\caption{$g_{1exp.}^{n}(x,Q^2)$ of the E154 collaboration vs.~the 
theoretically corrected values for 
$g_{1}^{n}(x,Q^2)$, as given in  Eq.\ (\ref{master2}), as functions of $x$ at $Q^2$=5 GeV$^2$. 
The experimental data points and the corresponding statistical error bars are presented as solid dots and solid vertical lines, correspondingly. The theoretically corrected values for $g_{1}^{n}(x,Q^2)$ are given by open circles. (The open circles have been shifted to make them legible.)}
\label{fig:three}
\end{figure}

\begin{figure}
\hspace{-2cm}
\includegraphics[scale=.9]{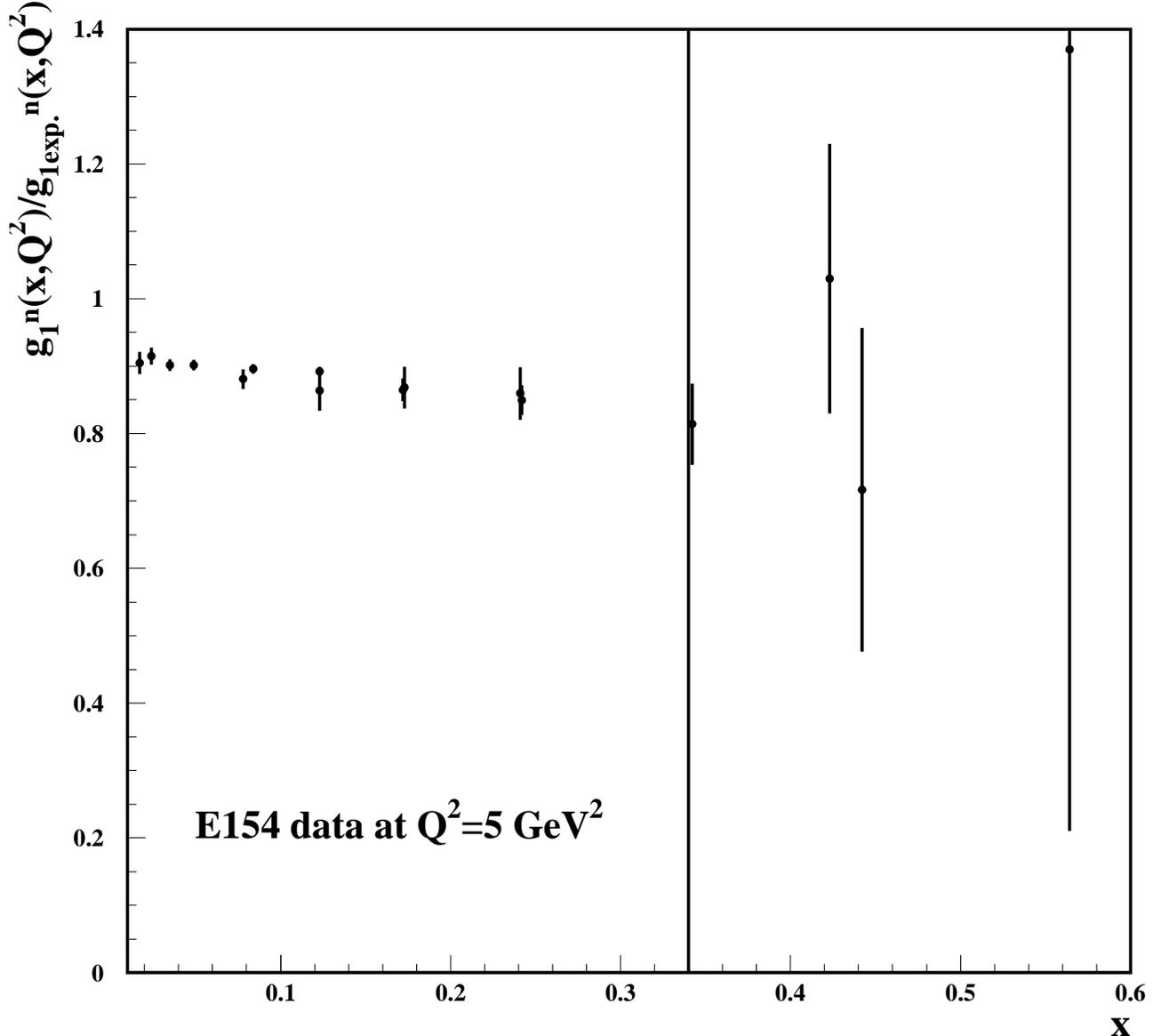}
\caption{The ratio $g_{1}^{n}(x,Q^2)/g_{1exp.}^{n}(x,Q^2)$ of Eq.\ (\ref{master3})  as a function of $x$ at $Q^2$=5 GeV$^2$ for the E154 data.   
The vertical solid lines are systematic uncertainties of experimental values for  $g_{1}^{n}(x,Q^2)$ and $g_{1}^{p}(x,Q^2)$ added in quadrature.
}
\label{fig:four}
\end{figure}

\begin{figure}
\hspace{-2cm}
\includegraphics[scale=.9]{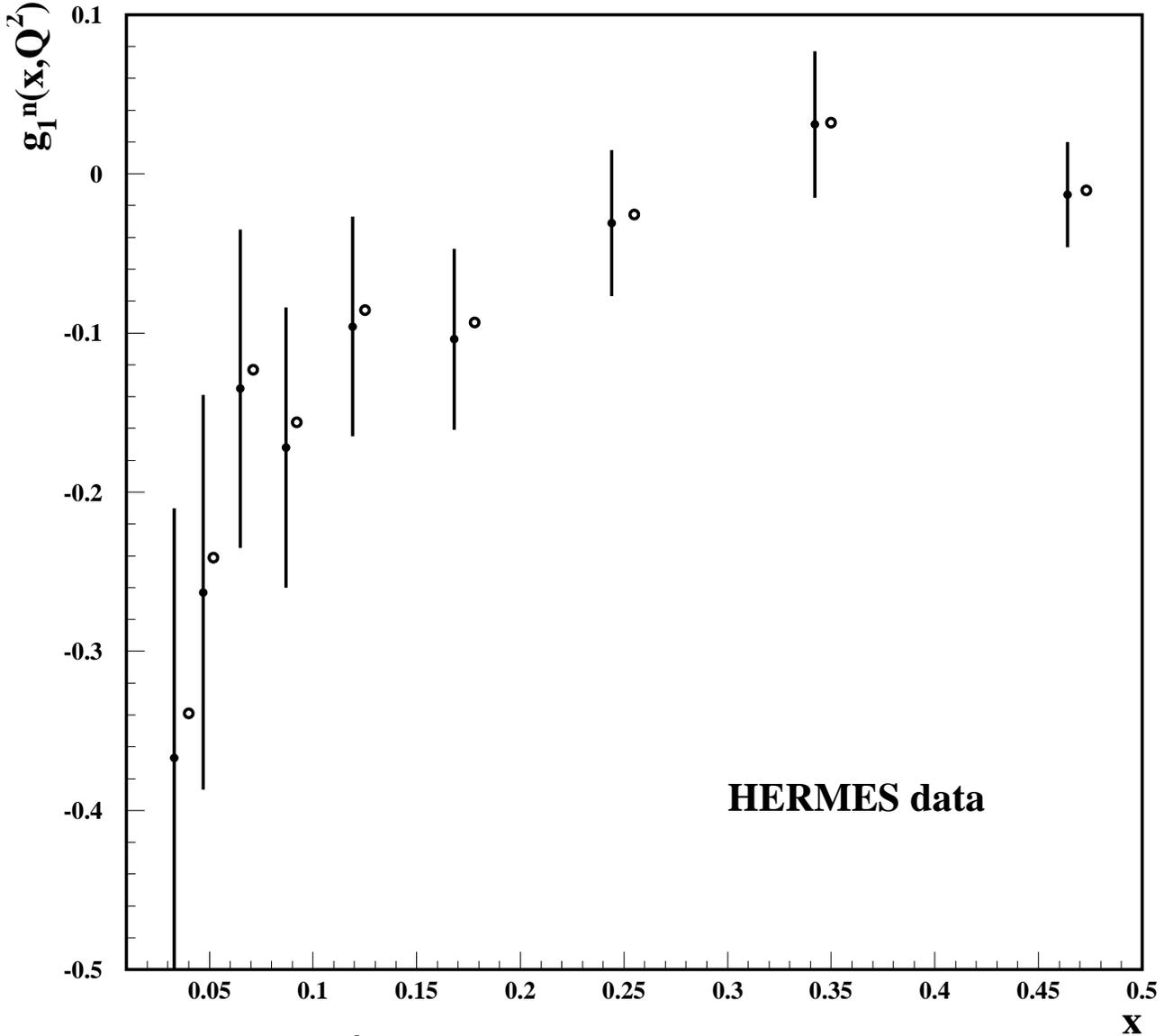}
\caption{
$g_{1exp}^{n}(x,Q^2)$ of the HERMES collaboration vs.~the theoretically corrected values for $g_{1}^{n}(x,Q^2)$, as given in Eq.\ (\ref{master2}),  as functions of $x$ at $Q^2$ correlated with $x$. 
The experimental data points and the corresponding statistical error bars are presented as solid dots and solid vertical lines, correspondingly. The theoretically corrected values for $g_{1}^{n}(x,Q^2)$ are given by open circles. (The open circles have been shifted to make them legible.)
}
\label{fig:five}
\end{figure}

\begin{figure}
\hspace{-2cm}
\includegraphics[scale=.9]{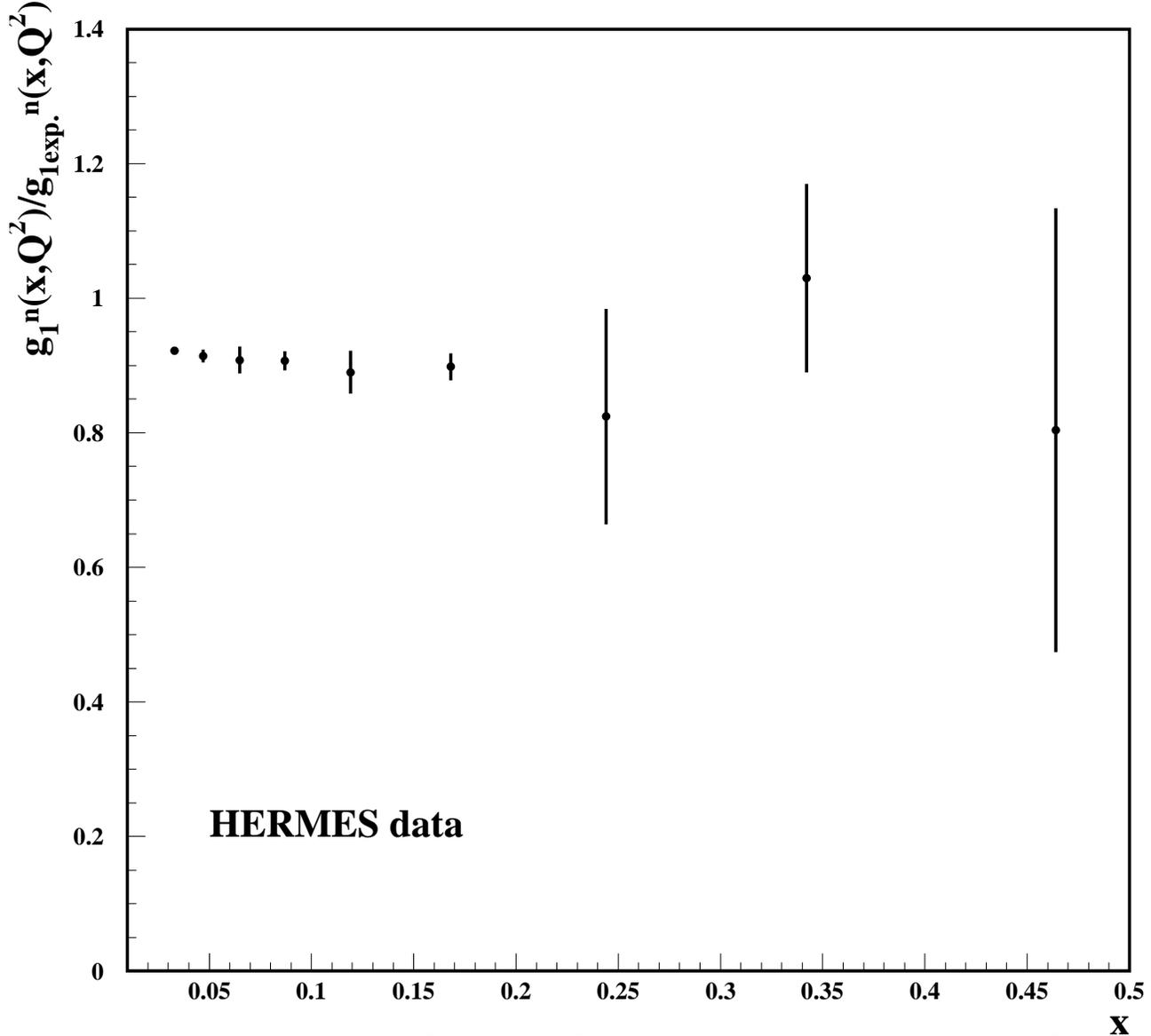}
\caption{The ratio $g_{1}^{n}(x,Q^2)/g_{1exp.}^{n}(x,Q^2)$ of Eq.\ (\ref{master3})  as a function of $x$ and at $Q^2$, correlated with $x$ for the  HERMES data.
The vertical solid lines are systematic uncertainties of experimental values for  $g_{1}^{n}(x,Q^2)$ and $g_{1}^{p}(x,Q^2)$ added in quadrature.
}
\label{fig:six}
\end{figure}

 Eqs.\ (\ref{master2}) and (\ref{master3}) can be used in order to re-analyze the extraction of the neutron spin structure function $g_{1}^{n}(x,Q^2)$ from the $^3$He DIS data. The results are presented in Figs.\ \ref{fig:three}, \ref{fig:four}, \ref{fig:five} and \ref{fig:six}.

Figure \ref{fig:three} represents the values of $g_{1exp.}^{n}(x,Q^2)$ reported by  the E154 collaboration \cite{E154} as solid dots, with corresponding error bars, and the theoretical prediction for $g_{1}^{n}(x,Q^2)$ of Eq.\ (\ref{master})  as open circles. (The open circles have been shifted to the right in order to make them legible on the plot.) The  error bars are purely statistical uncertainties and $Q^2$=5 GeV$^2$  for each point. The values of $g_{1}^{p}(x,Q^2)$ needed for Eq.\ (\ref{master2}), at appropriate values of $x$ and $Q^2$=5 GeV$^2$, were taken from the data of the SLAC E143 experiment \cite{E143}. From Fig.\ \ref{fig:three} one can see that the effect of the  $\Delta$ on the extraction of $g_{1}^{n}(x,Q^2)$ from the $^3$He data is to increase the values of $g_{1}^{n}(x,Q^2)$ by $12 \div 20$\% in the range $0.05 \le x \le 0.34$, which is still within the experimental error bars. At $x > 0.34$ the correction due to the $\Delta$ is of order $20 \div 40$\%. However,  the error bars for the corresponding values of  $g_{1exp.}^{n}(x,Q^2)$ are so large that it does not seem sensible to discuss any comparison with the data for $x > 0.34$, at present.
 Note, however, that high precision  measurements with the polarized $^3$He target are planned at TJNAF for the $0.33 \le x \le 0.63$ region \cite{Meziani}.

The information in Fig.\ \ref{fig:three} is presented in terms of the ratio of  Eq.\ (\ref{master3}) in  Fig.\ \ref{fig:four}. The error bars are systematic uncertainties of experimental values for  $g_{1}^{n}(x,Q^2)$ and $g_{1}^{p}(x,Q^2)$ added in quadrature.
  
In Figs.\ \ref{fig:five} and \ref{fig:six} we analogously re-analyze the HERMES data Ref. \cite{HERMES}. Note that the data points in  Ref. \cite{HERMES} have not been evolved to a common $Q^2$. Thus, in Figs.\ \ref{fig:five} and \ref{fig:six} the  values of $Q^2$ and $x$ are correlated.
We used the HERMES proton data \cite{HERMESp} for  $g_{1}^{p}(x,Q^2)$ in order to use in Eq.\ (\ref{master2}) since the values of $\langle x \rangle$ and $\langle Q^2 \rangle$ presented are very close to those of Ref.\ \cite{HERMES}.
 In order to have $g_{1}^{p}(x,Q^2)$ at exactly the same bins in $\langle Q^2 \rangle$ as for $g_{1}^{n}(x,Q^2)$ we have extrapolated $g_{1}^{p}(x,Q^2)$ to the required $\langle Q^2 \rangle$ using the experimentally justified assumption that the ratio $g_{1}^{p}(x,Q^2) / F_{1}^{p}(x,Q^2)$ is $Q^2$-independent \cite{E155}. The spin-independent structure function $F_{1}^{p}(x,Q^2)$ was parameterized using the recent world averaged fits for $R(x,Q^2)$ \cite{R} and $F_{2}^{p}(x,Q^2)$ \cite{F2}. One can see from Figs.\ \ref{fig:five} and \ref{fig:six} that the contribution of the interference terms increases $g_{1}^{n}(x,Q^2)$ by $10 \div 25$\%, which is so far within the experimental error bars.

\section{Conclusions}

In this work we considered a novel nuclear contribution which affects the extraction of the neutron spin structure function $g_{1}^{n}(x,Q^2)$ from the polarized DIS data on $^3$He.
The Feynman diagrams which describe  deep inelastic scattering on polarized $^3$He are  analogous to the diagrams which enter the calculation of the Gamow-Teller matrix element of the tritium beta decay. Thus, it is very natural to assume that the diagrams associated with the transitions $n \rightarrow \Delta^{0}$ and  $p \rightarrow \Delta^{+}$  play an important role in polarized DIS on $^3$He, because two-body exchange currents involving $\Delta$ isobars are important in calculations of the Gamow-Teller matrix element. 

The contributions of the interference transitions $n \rightarrow \Delta^{0}$ and  $p \rightarrow \Delta^{+}$ to the spin structure function of $^3$He, $g_{1}^{^3He}(x,Q^2)$, are characterized by the interference spin structure functions $g_{1}^{n \rightarrow \Delta^{0}}(x,Q^2)$ and $g_{1}^{p \rightarrow \Delta^{+}}(x,Q^2)$, and effective polarizations $P_{n \rightarrow \Delta^{0}}$ and $P_{p \rightarrow \Delta^{+}}$, respectively. A new sum rule for $g_{1}^{n \rightarrow \Delta^{0}}(x,Q^2)$ and $g_{1}^{p \rightarrow \Delta^{+}}(x,Q^2)$ has been derived.
We also related  $g_{1}^{n \rightarrow \Delta^{0}}(x,Q^2)$ and $g_{1}^{p \rightarrow \Delta^{+}}(x,Q^2)$ to $g_{1}^{n}(x,Q^2)$ and $g_{1}^{p}(x,Q^2)$ within the framework of the valence quark model. 
The connection 
to the calculations of the triton beta decay enabled an estimate of the effective polarizations of the interference terms, $P_{n \rightarrow \Delta^{0}}$ and $P_{p \rightarrow \Delta^{+}}$, 
 by requiring that the generalization of the Bjorken sum rule to the tri-nucleon system be consistent with the measured axial  coupling constant in the $A=3$ system.

Taking the effect of the $\Delta$ into account, we have re-analyzed the neutron spin structure function $g_{1}^{n}(x,Q^2)$ using the data of the E143 and HERMES experiments. We found that, depending on $x$, the values of $g_{1}^{n}(x,Q^2)$ increase by $10 \div 40$\%. We also estimated that the first moment of $g_{1}^{n}(x,Q^2)$ increases by $6 \div 8$\%.

\section{Acknowledgements}

V.G. would like to thank Kazuo Tsushima for many useful discussions and pointing to Refs.\ \cite{Alvarez} and \cite{Kitagaki}. This work was partially supported by the Australian Research Counsel and the Department of Energy of the United States of America.

\begin{figure}
\hspace{-2cm}
\includegraphics[scale=1.1]{rr1.epsi}
\vskip 2 cm
\caption{This figure demonstrates the correspondence between the Feynman diagrams describing the two-body exchange currents involving the $\Delta$ isobar which appear in calculations of the triton beta decay and the diagrams involving the  $n \rightarrow \Delta^{0}$ and $p \rightarrow \Delta^{+}$ transitions which contribute to the polarized DIS on $^3$He.
}
\label{fig:one}
\end{figure}

\end{document}